\documentclass{article}
\usepackage[T1]{fontenc}
\usepackage[utf8]{inputenc}
\usepackage[]{ismir} 
\usepackage{amsmath,cite,url}
\usepackage{graphicx}
\usepackage{color}
\newcommand{\datasetname}{BAMM}
\usepackage{booktabs}

\title{Assessing AI-generated music detection in real-world broadcast monitoring}

\multauthor
  {David López-Ayala$^1$ \hspace{1cm} Fernando Garcia de la Cruz$^1$ \hspace{1cm} Pablo Zinemanas$^2$}
  {{\bf Emilio Molina$^2$ \hspace{1cm} Martín Rocamora$^1$ }\\
  $^1$ Music Technology Group, Universitat Pompeu Fabra, Barcelona, Spain\\
  $^2$ BMAT Licensing S.L., Barcelona, Spain\\
  {\tt\small jorgedavid.lopez@upf.edu}
  }

\def\authorname{D. López-Ayala, F. Garcia de la Cruz, P Zinemanas, E Molina, and M. Rocamora}

\begin{document}

\maketitle

\begin{abstract}
The proliferation of AI-generated music in broadcast media raises concerns about transparency and fair compensation, but reliable detection under real broadcast conditions remains unresolved. Existing studies report substantial performance degradation in this domain, yet their evaluations are limited to synthetic broadcast data. To address this gap, we introduce \datasetname{} (Broadcast AI-Music Monitoring), a 40-hour dataset of real-world television recordings containing AI-generated and human-made music. We compare clean-trained and broadcast-trained CNN variants across three progressively more challenging scenarios: Clean Foreground Music (CFM), Synthetic TV Broadcast (STB), and Real TV Broadcast (RTB). Both models achieve near-perfect performance on CFM but degrade substantially under synthetic broadcast conditions. Broadcast-oriented training improves robustness compared with clean training, although performance remains limited. On RTB, evaluated using \datasetname{}, both models degrade further and show substantial score overlap between AI-generated and human-made music. These results expose a critical domain gap and show that current training approaches on CNN-based detectors remain insufficient for reliable AI-generated music detection in broadcast monitoring.
\end{abstract}

\section{Introduction}\label{sec:introduction}

Generative music models have rapidly lowered the barrier to music production. Their widespread adoption has brought to light a range of technical and legal challenges concerning training data provenance, training data replication, the definition of authorship, and compensation frameworks. These tensions have resulted in lawsuits against AI-music companies such as Suno and Udio \cite{tencer2024MajorRecordCompanies}, as well as licensing deals between generative-music platforms and rightsholders \cite{stassen2025WarnerMusicGroup}. In parallel, policies such as the EU AI Act introduce transparency requirements for AI-generated content to make AI-generated media identifiable \cite{serra2025marking}.

The need for robust detection tools has become increasingly urgent as synthetic content reaches an industrial scale. This urgency is underscored by Deezer’s report that, for the first time, 50\% of its daily uploads were detected as fully AI-generated.\cite{wendel2026AIMusicTops}. Similar trends have also been observed in broadcast media, with companies such as BMAT reporting a growing presence of AI-generated music in television recordings. \cite{bolboaca2026FirstVideoKilled}.

\begin{figure} 
\centering 
\includegraphics[alt={General Diagram},width=1\linewidth]{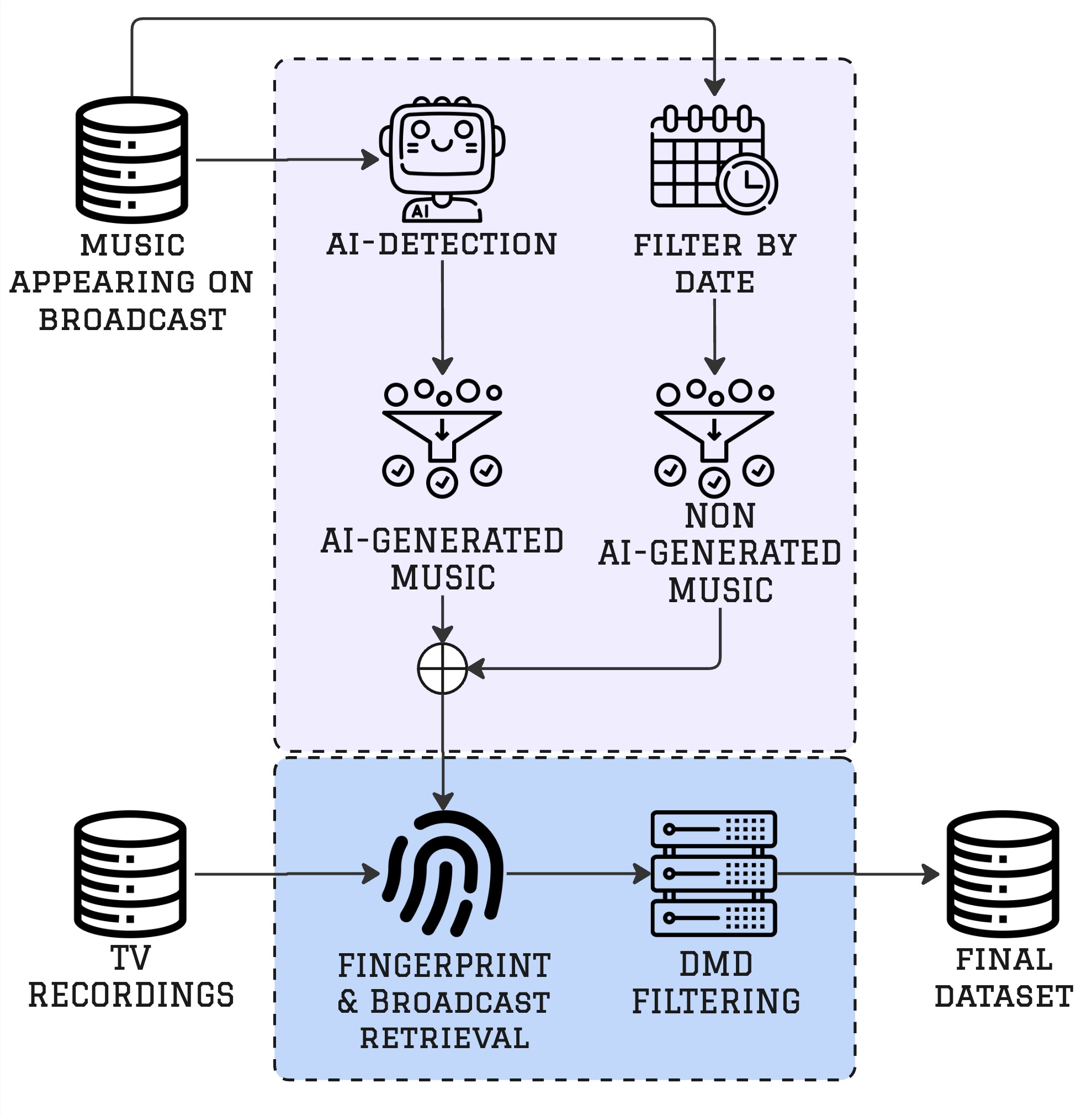} 
\caption{Generation pipeline for \datasetname{} dataset.} 
\label{fig:dataset_pipeline} 
\end{figure}

While existing AI-generated music detectors achieve near-perfect accuracy on isolated, high-fidelity tracks, they are often not robust to domain shifts and common audio transformations, such as random pitch shifting, time stretching, EQ, reverb, among others \cite{afchar2025AIGeneratedMusicDetectiona}. The broadcast environment introduces significant acoustic complexities: music segments are typically short, frequently backgrounded by dominant speech or sound effects, and subjected to transmission constraints. Recent benchmarks have demonstrated that, even in synthetic mixtures, these factors substantially degrade performance on state-of-the-art architectures \cite{loay2026aiopenbmat}. However, the extent to which AI detectors work in real-world broadcast conditions remains under-explored. 

To bridge this gap, we introduce \datasetname{}, a 40-hour dataset of real TV broadcast content specifically curated for AI-generated music detection. The dataset is balanced, containing approximately 20 hours of AI-generated music and 20 hours of human-made music, and was built by matching clean references to their broadcast occurrences through audio fingerprinting. The dataset contains clips between 5~s and 60~s, with music appearing either in the foreground or background. By collecting material from real TV content aired worldwide, \datasetname{} captures the variability and practical constraints of broadcast environments. We leverage this dataset to evaluate state-of-the-art CNN models, and identify current architectural limitations, and thus provide a reference evaluation setting for real-world AI music identification. The \texttt{.mp4} recordings are publicly available on Zenodo, and the baseline code and evaluation scripts are in Github \footnote{https://github.com/DaveLoay/BAMM-dataset.git}.

\section{Related Work}

The detection of AI-generated music has emerged within the MIR field as part of a broader effort to identify synthetic audio, alongside related tasks such as voice spoofing and synthetic speech detection
\cite{li2024AudioDeepfakeDetection, sing_deepfake}. Music-specific work has primarily focused on unintentionally left artifacts by generative systems. Afchar et al.\ \cite{afchar2025AIGeneratedMusicDetectiona} showed that these artifacts introduced during the decoding stage can be detected, under controlled conditions, with CNN-based models, using neural audio codecs such as Encodec and DAC \cite{defossez2022HighFidelityNeural,kumar2023high} to simulate this process. In later work, Afchar et al.\ \cite{afchar2025FourierExplanationAImusic} further characterized these artifacts through Fourier theory and showed that similar patterns are also present in music generated by commercial platforms such as Suno and Udio.

Then Rahman et al.\ \cite{rahman2024sonics} introduced SONICS, a dataset composed of songs generated with Suno and Udio for the AI-generated class and samples from the Genius Lyrics Dataset \cite{gdcjGeniusSongLyrics} for the human-made class. They also proposed the \textit{SpecTTTra} family of models, demonstrating that spatio-temporal representations are effective for distinguishing AI-generated from human-made music, achieving near-perfect performance. However, despite promising results, existing approaches remain largely limited to controlled environments. Cros Vila et al. \cite{vila2025AIMusicArms} showed that variations in sampling rate and bit rate may encourage detectors to rely on dataset-specific shortcuts rather than features intrinsic to AI-generated music, while models with near-perfect in-domain performance may generalize poorly beyond their training conditions.

These works establish strong baselines, constraints, and limitations for the task, showcasing that robust AI-generated music detection remains unresolved in more complex scenarios. One such scenario is broadcast, which introduces substantially different conditions: music may appear in the foreground or background and may be mixed with speech and sound effects at different relative loudness levels. Previous work has addressed these challenges through distinct tasks. Relative music loudness estimation aims to characterize how music appears within a broadcast mix \cite{melendezcatalan2021RelativeMusicLoudness}, with OpenBMAT providing a dataset specifically designed for this purpose \cite{melendez-catalan2019OpenBroadcastMedia}. Audio fingerprinting, in contrast, focuses on matching reference tracks against TV recordings \cite{cortes-sebastia2025EnhancedTelevisionBroadcast}, with BAF providing a benchmark for evaluating this task under broadcast conditions \cite{cortes2022BAFAudioFingerprinting}. Together, these works highlight the prevalence of background music and mixed-audio conditions in real broadcast content.

López-Ayala et al.\ \cite{loay2026aiopenbmat} addressed AI-generated music detection in broadcast settings using synthetically constructed television content. Their study showed that existing detectors experience substantial performance degradation when music appears in short excerpts, is masked by dominant speech or sound effects, or is subject to broadcast-specific production and storage constraints. These include transitions, editing practices, changes in dynamics, and low-quality audio encoding, as further discussed in Section~\ref{sec:evaluation-scenarios}. To support this evaluation, they introduced AI-OpenBMAT, a dataset designed for AI-generated music detection in broadcast settings. It reproduces the duration patterns and loudness relationships of real television audio by combining human-made production music with stylistically matched continuations generated using Suno. Although AI-OpenBMAT provides a valuable controlled benchmark, its broadcast conditions are still simulated through synthetic mixtures. This work extends that line of research by introducing a dataset built from real television broadcast recordings and using it to evaluate current AI-generated music detectors under practical broadcast conditions.

\section{\datasetname{} DATASET}\label{sec:DATASET}

\datasetname{} (Broadcast AI-Music Monitoring) is a novel dataset comprising 40 hours of real-world broadcast content curated for AI-generated music detection. The dataset was constructed by identifying clean AI-generated and human-made reference tracks within a global broadcast archive using audio fingerprinting, followed by a multi-stage filtering process to retain valid broadcast occurrences. BAMM consists of approximately 20 hours of AI-generated music and 20 hours of human-made music, stored as \texttt{.mp4} clips. Each sample captures original broadcast audio and video, preserving the authentic acoustic degradations and production contexts (e.g., music in the background, sound effects) found in practice. The clips were sampled from a global broadcast archive that monitors over 4{,}200 television channels worldwide, with all samples aired between January 2025 and March 2026 and durations ranging from 5~s to 60~s. Reflecting the technical constraints of large-scale industrial monitoring, the audio is monophonic, sampled at 8~kHz, and encoded with AAC-LC at bitrates of at least 40~kbps. While lower than consumer broadcast standards, these specifications are representative of the low-bitrate proxy streams used in global industrial monitoring, posing a significant worst-case challenge for detection models.

To ensure trustworthy ground truth labels, we developed an automated pipeline to bridge the gap between detecting AI content in clean conditions and real-world broadcast. We first curated a reference corpus of clean, foreground \texttt{.mp3} tracks for both the AI-generated and human-made classes. We then leveraged an audio fingerprinting system to identify the precise occurrences of these tracks within the global broadcast archive. The retrieved segments were filtered using a multi-stage process devised to isolate valid clips that satisfied the technical requirements described above, as illustrated in Fig.~\ref{fig:dataset_pipeline}. The following subsections describe the individual stages of this procedure.

\subsection{Reference Track Selection}

The construction pipeline begins with a candidate pool of clean MP3 reference tracks. Before searching for occurrences of these tracks in TV broadcast content, we categorize them using an ensemble of five AI-music detectors, hereafter referred to as the \textit{detector ensemble}. This configuration promotes methodological diversity by combining detectors trained on different datasets, operating at different sampling rates, and relying on distinct feature representations.

The labels were determined separately for both classes. For the human-made class, reference tracks were restricted to releases between January 2020 and December 2022, ensuring that they predated Suno v3.5 and other publicly available or commercial generative music systems capable of producing music of comparable quality. For the AI-generated class, a candidate track was included only if it exceeded the calibrated decision thresholds across all five detectors, requiring unanimous agreement from the \textit{ensemble}.

 Regarding the models in the \textit{detector ensemble}. The first model is a publicly available convolutional neural network presented and assessed in this paper under the name \textit{CNN Clean}. It is based on the architecture proposed by Afchar et al. [6], was trained on publicly available data, It operates on 5-second audio windows represented as mel-spectrograms computed from audio resampled to 8~kHz. The model weights and inference code are available in the project repository. For more in-depth implementation details, refer to Section~4.1.

The second model is the publicly available \textit{SpecTTTra} alpha-5s detector \cite{rahman2024sonics}. It was selected because it achieved the best performance among the evaluated \textit{SpecTTTra} variants on our control dataset. The model operates on spatio-temporal tokens extracted from audio resampled to 16~kHz and was trained on the SONICS suno v3.5 subset.

The remaining three models of the \textit{detector ensemble} were developed for internal research purposes and trained on in-house data. The third model is a CNN based on the architecture proposed by Afchar et al. \cite{afchar2025AIGeneratedMusicDetectiona}. It operates on 5-second audio windows represented as mel-spectrograms computed from audio resampled to 16~kHz. The fourth model is a logistic regression classifier following the approach introduced by Afchar et al. \cite{afchar2025FourierExplanationAImusic}. It operates on artifact-fingerprint features extracted from audio resampled to 16~kHz. The fifth model is an MLP trained on the same feature representation as the logistic regression classifier.

All five models were trained as binary classifiers under clean foreground music conditions to distinguish human-made from AI-generated tracks (Suno v3.5), and all achieved F1 scores above 98\%. We performed a conservative calibration on each model to achieve a zero false positive rate. The calibration set included 5,000 tracks from Da-TACOS \cite{yesiler2019}, providing a verified human baseline that predates modern generative models, and a control set of 500 AI-generated tracks from the clean subset of AI-OpenBMAT. By deriving model-specific decision thresholds that yielded zero errors on the human-composed set, we obtained a labeling criterion with high confidence in the resulting AI-generated samples. 

\begin{figure}[t]
    \centering
    \includegraphics[width=1\linewidth]{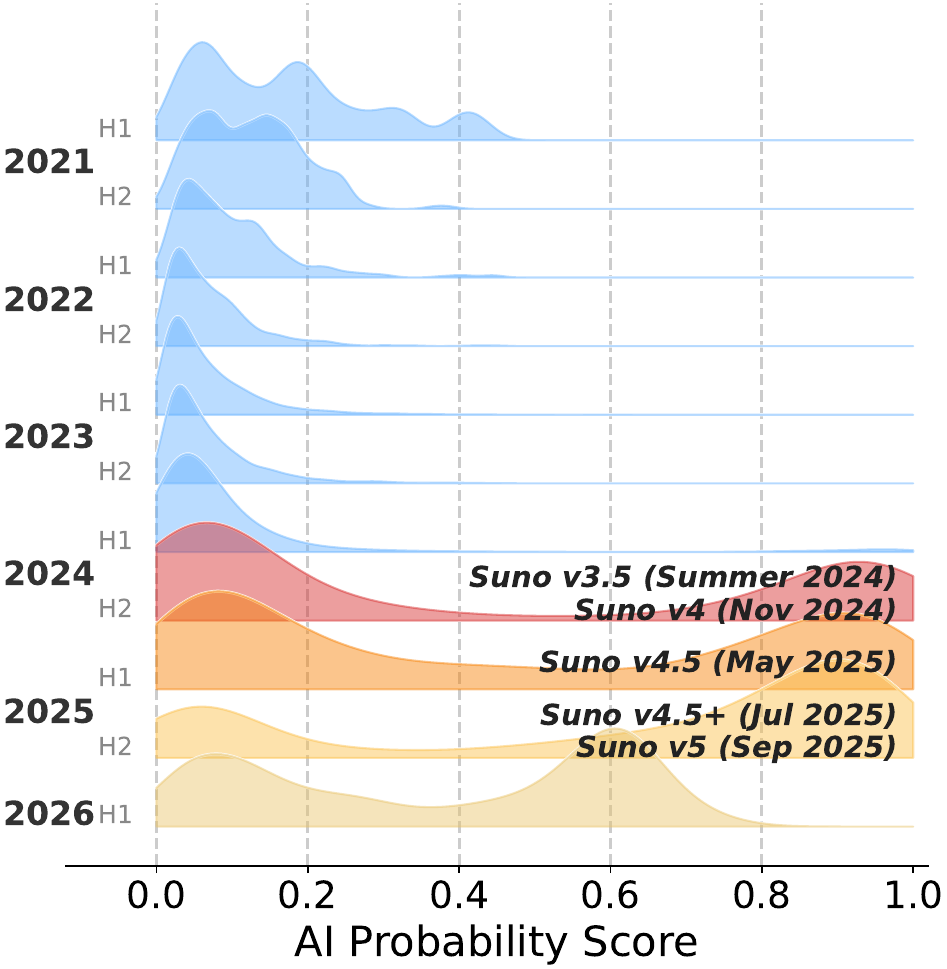}
    \caption{Distribution of AI-probability scores produced by the \textit{detector ensemble} for tracks released from 2021 to 2026. Score values remain low before the release of Suno v3.5, and increase afterward, indicating the appearance of Suno-generated music.}
    \label{fig:ai-over-time}
\end{figure}

To further validate the selection strategy, we analyzed the temporal evolution of AI-positive detections among the artists with the highest numbers of positive cases, as shown in Figure \ref{fig:ai-over-time}. Detection scores are consistently low before the Summer 2024 release date of Suno v3.5, increase markedly after its introduction, and later decrease over time. This behavior is consistent with the real-world adoption of Suno v3.5 and the subsequent appearance of newer generative models from the same family.

\subsection{Fingerprint \& Broadcast Retrieval}

Once the reference tracks for AI-generated and human-made classes were defined, we used a private implementation of audio fingerprinting based on landmarks \cite{cortessebastia2025MusicIdentificationAudio} to locate matches in our TV broadcast archive. The matched segments were searched in recordings aired between January 2025 and March 2026 and then extracted directly from the original TV emissions. These recordings are archived at a sampling rate of 8~kHz and a bitrate of 40~kbps.

\subsection{DMD Filtering}
Following fingerprint retrieval, we applied a Deep Music Detector (DMD), based on the approach proposed by Meléndez et al.\ \cite{melendezcatalan2021RelativeMusicLoudness}. This model classifies audio segments as foreground music, background music, or speech. We kept only clips labeled as containing music, whether in the foreground or background, thereby removing sound effects and incorrect fingerprint matches. These foreground and background labels are also used when analyzing the performance of AI-generated music detection models (Section~\ref{sec:ResultsRTB}). 

\section{Experimental Setup}\label{sec:EXPERIMENTAL-SETUP}

This benchmark examines how the training domain affects CNN-based detectors designed to identify AI-generated music in real broadcast settings. We compare two variants of the architecture proposed by Afchar et al.\ \cite{afchar2025AIGeneratedMusicDetectiona}, both operating on 5-second windows but trained under different conditions: clean foreground music and broadcast-oriented data. This comparison isolates the effect of the training domain while keeping the model capacity fixed, allowing us to examine the limitations of CNN-based AI-generated music detectors in broadcast environments. The benchmark focuses on Suno v3.5, as model-agnostic detection remains an open problem, and this generative model has publicly available data and matched reference detectors.

\subsection{Shared Architecture and Preprocessing}
The CNN-based model has six convolutional layers with filter sizes $[16, 32, 64, 128, 256, 512]$ and kernel size 3, followed by average pooling and two fully connected layers.

In the pre-processing pipeline, audio is converted to mono and resampled to 8~kHz. Then the waveform is transformed into a time-frequency representation using an STFT with a window size of 2048 and a hop length of 1024. The resulting power spectrogram is projected onto 128 Mel bands spanning 20~Hz to 4~kHz, converted to the logarithmic dB scale, and standardized with a fixed global mean of $-4.0$ and standard deviation of $3.0$. The Mel-spectrogram is then segmented into discrete time windows to form the input tensor of the CNN.

\subsubsection{CNN Clean}

\textit{CNN Clean} is trained on clean foreground music only. Audio is converted to mono and resampled to 8~kHz in the pre-processing pipeline. The human-made class is drawn from the FMA-medium dataset, comprising approximately 25k songs, while the AI-generated class is built from the Suno v3.5 subset of SONICS, comprising approximately 19k songs. For each batch, five 5~s snippets are randomly sampled from each track, enabling consistent training.

\subsubsection{CNN Broadcast}

\textit{CNN Broadcast} uses the same architecture and input representation as \textit{CNN Clean}, but is trained on data designed to emulate broadcast conditions by mixing music and speech. The music sources are the same as for \textit{CNN Clean}, namely FMA-medium for the human-made class and the Suno v3.5 subset of SONICS for the AI-generated class. Speech is taken from the 360-hour clean training partition of LibriSpeech \cite{korvas_2014}.

The training set is constructed with a controlled ratio of 70\% mixed samples and 30\% clean samples for each class. In the mixed condition, speech segments are concatenated to span the full duration of each music track and mixed with the music signal at a randomly sampled SNR between $-30$~dB and $+30$~dB. The resulting files are exported in mono at 8~kHz and encoded as \texttt{.mp3} at 40~kbps before entering the shared preprocessing pipeline.

\subsection{Evaluation Scenarios}\label{sec:evaluation-scenarios}

To assess the benchmark models under progressively more challenging conditions, we evaluate them across three scenarios: (i) Clean Foreground Music (CFM), (ii) Synthetic TV Broadcast (STB), and (iii) Real TV Broadcast (RTB).

\subsubsection{Clean Foreground Music (CFM).}

For this scenario, we evaluate the models on the clean music tracks used to build AI-OpenBMAT \cite{loay2026aiopenbmat}. The dataset contains 476 human--AI pairs with a sampling rate of 22.05~kHz and a bitrate of 353~kbps. The AI tracks were generated from their corresponding human tracks using Suno's \textit{extend} function. As a result, the pairs are closely matched in musical characteristics such as style, timbre, pitch, and key. This scenario provides a direct point of comparison with the synthetic broadcast scenario, which is built from the same source material.

\subsubsection{Synthetic TV Broadcast (STB).}
We evaluate the synthetic broadcast scenario using the full AI-OpenBMAT dataset, following \cite{loay2026aiopenbmat}. The dataset contains 3{,}294 tracks, totaling 54.9 hours of synthetic broadcast material, with a sampling rate of 22.05~kHz and a bitrate of 353~kbps. These data were created from the same clean foreground references described above, using controlled mixtures designed to accurately emulate broadcast conditions. This scenario allows us to evaluate model performance under synthetic broadcast conditions while keeping the underlying musical material controlled.

\subsubsection{Real TV Broadcast (RTB).}
The real broadcast scenario uses the full \datasetname{} dataset introduced in this work. It comprises 40 hours of real TV broadcast recordings, with approximately 20 hours per class, at a sampling rate of 8~kHz and a bitrate of 40~kbps using AAC-LC encoding. This setting extends previous broadcast-oriented evaluations by shifting from synthetic approximations to real television content, enabling the models to be tested under the conditions of the target environment.

\begin{figure*}[t]
    \centering
    \includegraphics[width=\textwidth]{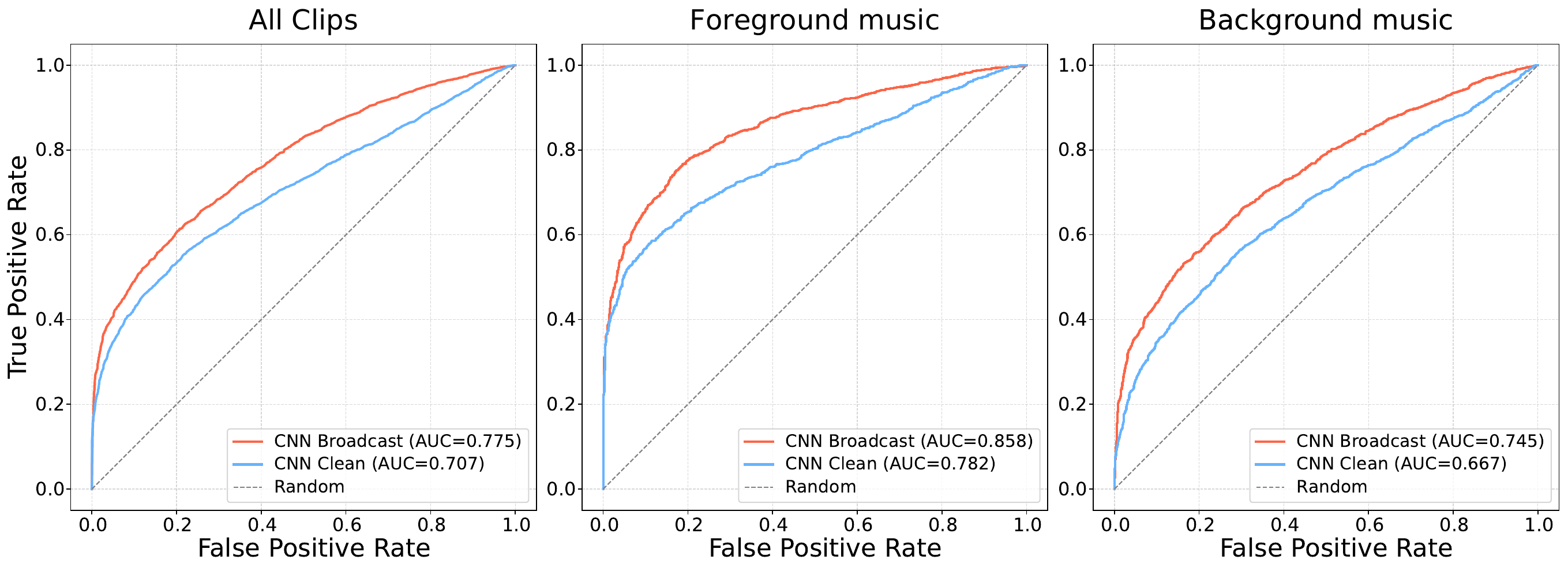}
\caption{ROC curves for the clean-trained and broadcast-trained CNN models when evaluated on all BAMM clips, only on foreground music, and only on background music. Performance decreases for background music, showing that detection becomes more difficult when the musical signal is less salient or mixed with other content.}
\label{fig:roc-auc-curves}
\end{figure*}

\section{Results}\label{sec:Results}

We evaluate both models across three scenarios with increasing levels of broadcast complexity: Clean Foreground Music (CFM), Synthetic TV Broadcast (STB), and Real TV Broadcast (RTB). The comparison between \textit{CNN Clean} and \textit{CNN Broadcast} allows us to analyze how the training domain influences the robustness of AI-generated music detection. While the CFM scenario provides a controlled evaluation of artifact detection, the STB setup introduces synthetic broadcast conditions through controlled music--speech mixtures. Finally, the RTB case evaluates both models on the BAMM dataset introduced in this work, capturing the challenges of real television content.

\subsection{Evaluation on Clean Foreground Music}

The first stage of this evaluation aims to verify whether detectors trained under real-world broadcast constraints, i.e., low-quality audio at 8~kHz sampling rate, remain reliable for detecting AI-generated music under clean, high-quality foreground conditions. As shown in Table~\ref{tab:benchmark_results}, both \textit{CNN Clean} and \textit{CNN Broadcast} achieve F1-scores above 99\%, demonstrating that the artifacts learned from low-quality training data are still present in the high-quality data typically used for this task and remain detectable even when considering only the lower-frequency components of the audio spectrum. These results validate the proposed models and provide a baseline for the subsequent evaluations.

\begin{table}[t]
\centering
\small
\renewcommand{\arraystretch}{1.10}
\setlength{\tabcolsep}{4pt}
\begin{tabular*}{\columnwidth}{
@{\extracolsep{\fill}}
lcccc
@{}
}
\toprule
\textbf{Scenario} &
\multicolumn{2}{c}{\textbf{CNN Clean}} &
\multicolumn{2}{c}{\textbf{CNN Broadcast}} \\
\cmidrule(lr){2-3}
\cmidrule(lr){4-5}
& \textbf{F1} & \textbf{ROC} & \textbf{F1} & \textbf{ROC} \\
\midrule

CFM
& \textbf{0.992} & \textbf{0.998}
& \textbf{0.992} & 0.996 \\

STB
& 0.342 & 0.909
& \textbf{0.661} & \textbf{0.926} \\

RTB
& 0.186 & 0.707
& \textbf{0.472} & \textbf{0.775} \\

\bottomrule
\end{tabular*}
\caption{Results across the three evaluation scenarios: Clean Foreground Music (CFM), Synthetic TV Broadcast (STB), and Real TV Broadcast (RTB).}
\label{tab:benchmark_results}
\end{table}

\subsection{Evaluation on Synthetic TV Broadcast}

The synthetic TV broadcast (STB)  scenario represents the first evaluation under mixed-audio conditions, in which AI-generated music is no longer isolated but combined with speech. As observed in Table~\ref{tab:benchmark_results}, the transition from clean foreground music to synthetic broadcast conditions produces a significant reduction in detection performance for both models. The clean-trained model suffers the greatest degradation, demonstrating that artifacts learned from isolated music do not fully transfer when partially masked by the surrounding broadcast context. This result is consistent with the performance reported for AI-OpenBMAT \cite{loay2026aiopenbmat} and validates the experimental setup. The \textit{CNN Broadcast} model alleviates this degradation by learning from similar mixed conditions, demonstrating that exposure to broadcast-like data improves artifact detection under masking. However, its limited performance suggests that the problem is not only related to the training distribution but also to an intrinsic reduction in artifact visibility arising from the broadcast scenario's mixture process.

\subsection{Evaluation on Real TV Broadcast}\label{sec:ResultsRTB}

The RTB scenario is the most challenging setting in this work, as it evaluates the detectors directly on real television content. Unlike STB, which approximates broadcast conditions using controlled synthetic mixtures, RTB includes the full complexity of real recordings. As shown in Table~\ref{tab:benchmark_results}, both models perform worse than in STB, indicating that the synthetic setting does not fully capture real broadcast conditions.

The \textit{CNN Clean} model shows the largest degradation, confirming that artifacts learned from isolated foreground music do not transfer reliably to real broadcasts. In addition to the reduced audio quality of the RTB recordings (8~kHz sampling rate and 40~kbps AAC-LC), sound effects, transitions, and diverse mixing strategies can further obscure AI-related artifacts.

Figure~\ref{fig:roc-auc-curves} shows a detailed comparison of the models' performance when considering BAMM clips with music in the background or in the foreground. Both models retain some discriminative ability and perform better on foreground than on background music. This confirms that masking is a major source of degradation. However, the limited performance on foreground segments also indicates that degraded audio quality and other broadcast sounds affect artifact visibility.

\begin{figure}[ht]
    \centering
    \includegraphics[width=1\linewidth]{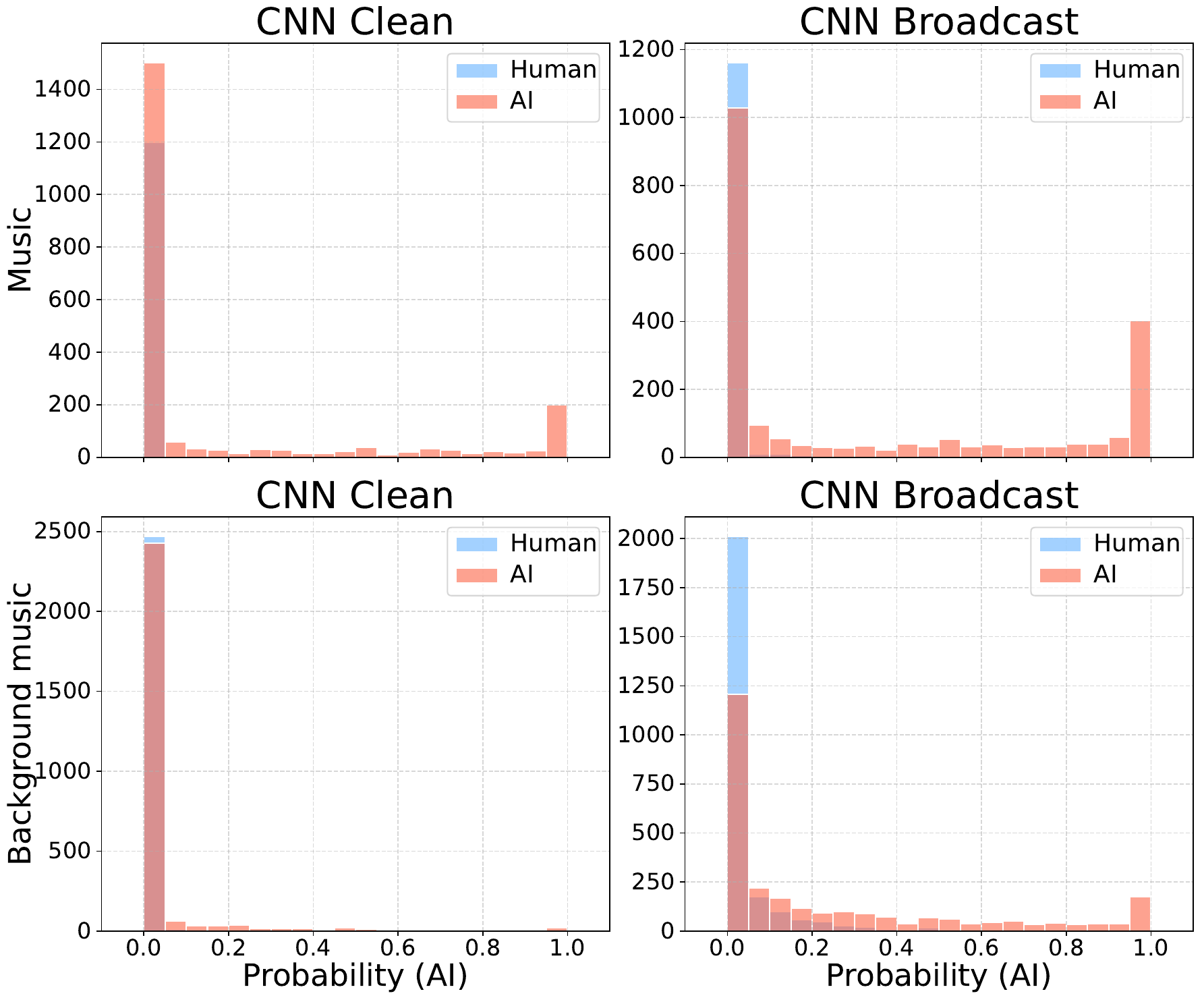}
   \caption{Score distributions for the CNN models under Real TV Broadcast conditions. Blue and red densities denote human-made and AI-generated clips.} 
    \label{fig:broadcast-hist}
\end{figure}

The score distributions in Figure~\ref{fig:broadcast-hist} show that the main difficulty is detecting AI-generated music rather than rejecting human content. While human samples are generally assigned scores near zero, many AI samples also receive low scores, particularly in background music. Overall, broadcast-simulated training improves robustness, but it does not fully overcome the effects of real broadcast conditions.

\section{CONCLUSIONS}\label{sec:CONCLUSIONS}

In this paper, we introduced \datasetname{}, a dataset for AI-generated music detection designed to address key limitations of existing benchmarks. Unlike datasets focused primarily on clean or isolated musical excerpts, \datasetname{} targets the real broadcast setting, where music appears in short segments, alternates between foreground and background roles,  and is mixed with speech and sound effects. 

Our experiments show that CNN-based detectors trained on clean foreground music or simulated broadcast data retain some discriminative capacity under controlled conditions, but their performance degrades substantially as the evaluation scenario becomes more realistic. In particular, although performance remains limited for both foreground and background music, the models perform slightly better when the music is in the foreground. This suggests that current detectors are not reliably capturing AI-generated music artifacts, and that the task becomes even more difficult when the musical signal is less salient or embedded within complex audio mixtures.

These findings highlight the need for more robust detection methods capable of modeling the acoustic and contextual complexity of real broadcast media. \datasetname{} provides a benchmark for this direction by exposing failure modes not fully captured by clean-music datasets and encouraging the development of models that can detect AI-generated music in challenging, real-world broadcast environments.

\section{ACKNOWLEDGMENTS}\label{sec:ACKNOWLEDGMENTS}

This work is supported by the "Cátedra IA y Música" project (TSI-100929-2023-1), funded by the Secretaría de Estado de Digitalización e Inteligencia Artificial, the European Union-Next Generation EU funds and BMAT Music Innovators. And by the "IMPA" project (PID2023-152250OB-I00) funded by MCIU/AEI/10.13039/501100011033/FEDER, UE.

\bibliography{ISMIRtemplate}

%
%
%
%

\end{document}